\documentclass{aa}
\usepackage{graphicx}
\usepackage{natbib}
\bibpunct{(}{)}{;}{a}{}{,}    
\usepackage[dvips]{color}
\newcommand{\beq}{\begin{equation}}
\newcommand{\eeq}{\end{equation}}
\newcommand{\bea}{\begin{eqnarray}}
\newcommand{\eea}{\end{eqnarray}}
\newcommand{\pdoverd}[2]{\frac{\partial #1}{\partial #2}}
\newcommand{\pdoverdt}[1]{\frac{\partial #1}{\partial t}}
\newcommand{\subscr}[1]{_\mathrm{#1}}

\newcommand{\Sect}[1]{Sect.~\ref{#1}}

\def\MSOL{{\rm M_\odot}}
\def\RSOL{{\rm R_\odot}}

\def\AU{{\rm AU}}
\def\PC{{\rm pc}}
\begin{document}
\title{Circumbinary Disk evolution}
\author{Richard G\"unther \& Wilhelm Kley}
\offprints{R. G\"unther,\\ \email{rguenth@tat.physik.uni-tuebingen.de}}
\institute{Institut f\"ur Astronomie \& Astrophysik, 
     Abt. Computational Physics,
           Auf der Morgenstelle 10, D-72076 T\"ubingen, Germany}
\date{Received January 21, 2002; accepted March 14, 2002}
\abstract{%
We study the evolution of circumbinary disks surrounding 
classical T Tau stars.
High resolution numerical simulations are employed to model
a system consisting of a central eccentric binary star within
an accretion disk. The disk is assumed to be infinitesimally thin,
however a detailed energy balance including viscous heating
and radiative cooling is applied.
A novel numerical approach using a parallelized Dual-Grid technique
on two different coordinate systems has been implemented.

Physical parameters of the setup are chosen to model
the close systems of \object{DQ Tau} and \object{AK Sco},
as well as the wider systems
of \object{GG Tau} and \object{UY Aur}. Our main findings are
for the tight binaries a substantial flow of material through the
disk gap which is accreted onto the central stars in
a phase dependent process. We are able to constrain the parameters
of the systems by matching both accretion rates and derived spectral
energy distributions to observational data where available.
}
\maketitle
\section{Introduction}
Observations of main sequence stars have shown that about 60\% belong
to binary or multiple systems \citep{1991A&A...248..485D}.
However, recent observations of pre-main sequence stars seem to indicate
that nearly all stars are born in multiple systems
\citep[Reviews in][]{2000prpl.conf..703M, 2001IAUS..200.....H}.

Apart from many spectroscopic binary systems with surrounding disks like
\object{DQ Tau} and \object{AK Sco},
there have been only two observations of directly imaged wide optical
binary systems with circumbinary disks, namely \object{GG Tau}
\citep{1994A&A...286..149D} and \object{UY Aur} \citep{1998A&A...332..867D}.
The observations of the spectroscopic binaries usually include emission
data from the infrared and optical part of the spectrum.
Observations of the luminosities of the boundary layer are typically used to
estimate accretion rates onto the binary stars
\citep{1998ApJ...492..323G, 1995ApJ...452..736H}.

\citet{1997MNRAS.285...33B} model the early phase of a molecular cloud
collapsing onto a protobinary system with an SPH code and derive
properties of the created circumbinary and circumstellar disks.
We are interested in the following phase of the evolution when the envelope
has been cleared.
By modeling the evolution we compare fully developed circumbinary and
circumstellar disks and their properties with observational data.

In a system consisting of a stellar binary surrounded by an
accretion disk there is a continuous exchange of angular momentum
between the binary and the disk. As the binary is orbiting with a
shorter period than the disk material it is generating positive
angular momentum waves in the disk which, upon dissipating, deposit
their energy and angular momentum in the disk. Hence, there is
a continuous transfer of angular momentum from the binary to the outer
disk. Upon receiving this angular momentum, the inner disk speeds up,
and the material is receding from the binary, leaving a region
of very low density around the central binary, a so called {\it inner gap}.
At the same time, the loss of angular momentum
leads to a secular shrinkage of the semi-major axis of the binary.
The width and form of the gap depend on disk parameters such as
temperature and viscosity and on the orbital parameters (eccentricity)
of the binary as well \citep{1991ApJ...370L..35A, 1994ApJ...421..651A}.

First calculations dealing with the evolution
of circumbinary disks so far have been using the numerical method of smoothed
particle hydrodynamics. Applying such a scheme 
\citet{1996ApJ...467L..77A} could demonstrate that for sufficiently
high viscosities and temperatures material from the disk is
able to penetrate into the inner gap region of the disk to become
eventually accreted by one of the binary stars. This process which is
facilitated by a gas flow along the saddle points of the potential
leads to a {\it pulsed accretion} whose 
magnitude and phase dependence for eccentric binaries
was estimated.
However, when using only a small number of particles all having
the same mass only a limited resolution is achievable. 

Later \citet{1997apro.conf..792R} used a finite
difference method in attacking this problem. Using a constant kinematic
viscosity coefficient and eccentric binaries they also found,
in agreement with \citet{1996ApJ...467L..77A}, a
{\it pulsed accretion flow} across the gap, with most of the
accretion occurring onto the lower mass secondary star.

In this paper we extend these computations and solve explicitly a
time dependent energy equation which includes the effects
of viscous heating and radiative cooling.
We aim at the structure and dynamics of the disk as well as
the gas flow in the close vicinity of the binary star.
To that purpose we utilize a newly developed method which enables us
to cover the whole spatial domain.
For the first time we perform long-time integration of the complete system
covering several hundred orbital periods of the binary and compare the
properties of the evolved systems with observational data such as
spectral energy distributions in the infrared and optical bands and
accretion rates estimated from luminosities.

In the next section we layout the physical model followed by
some remarks concerning important numerical issues (\Sect{sec:NM}). 
We describe the setup of the various numerical models in \Sect{sec:GMD}.
The main results are presented in \Sect{sec:results} and our 
conclusions are given in \Sect{sec:conclusion}. 
\section{Physical Model}
\label{sec:Phm}
\subsection{General Layout}
The system consists of a stellar binary system and a surrounding
circumbinary accretion disk.
We assume a geometrically thin disk and describe the 
disk structure by means of a two-dimensional, infinitesimal thin model
in the $z=0$ plane with the origin at the center of mass (COM) of the binary.
We use vertically averaged quantities, such as the surface mass density
\[ \Sigma = \int^\infty_{-\infty} \rho dz,\]
where $\rho$ is the regular three-dimensional density.
We work with two different coordinate systems.
Firstly, a cylindrical coordinate system $(r, \varphi)$ centered on
the COM of the binary stars. As such a coordinate system
does not allow for a full coverage of the plane, we overlay the center
additionally with a Cartesian $(x, y)$ grid (see Sec.~\ref{subsec:dgt}).

The gas in the disk is non-self-gravitating and is orbiting
a binary system with stellar masses $M_1$ and $M_2$ which
are modeled by the gravitational potential of two solid
spheres. The binary is fixed on a Keplerian orbit with given
semimajor axis $a$ and eccentricity $e$.
Hence, we neglect self-gravity and the gravitational backreaction
of the disk onto the stars. 
This is justified because the total mass of the disk $M_{\rm d}$
within the simulated region,
which extends up to several tens of semi-major axis $a$ of the binary,
is typically much smaller than that of the central binary. 
Hence, the timescale for a change
in $a$ and $e$ is indeed much longer than the timescales considered
here \citep{1991ApJ...370L..35A}.
For the majority of the computations we employ an inertial coordinate system, 
however test calculations were performed in a
system corotating with the binary as well.
\subsection{Equations}
The evolution of the disk is given by the two-dimensional
($x, y$ or $r, \varphi$) evolutionary equations for the
density $\Sigma$,
the velocity field $\vec{u}\equiv (u_x, u_y) = (u_r, u_\varphi)$,
and the temperature $T$.
In a coordinate-free representation the equations read
\beq
\label{eq:sigma}
 \pdoverdt{\Sigma} + \nabla \cdot \left( \Sigma {\vec{u}} \right) = 0
\eeq
\beq
\label{eq:momentum}
 \pdoverdt{\Sigma{\vec{u}}} 
   + \nabla\cdot\left(\Sigma \vec{u} \cdot \vec{u}\right) = 
   -\nabla P - \Sigma\nabla\Phi + \nabla\cdot{\bf T}
\eeq
\beq
\label{eq:temperature}
 \pdoverdt{\Sigma\varepsilon_{\rm tot}} 
 + \nabla \cdot \left[ \left( \Sigma\varepsilon_{\rm tot} 
       + P \right) {\vec{u}} \right] = - \Sigma \vec{u} \cdot \Phi
  - \nabla\cdot \left(\vec{F} - \vec{u} \cdot {\bf T} \right)
\eeq

Here $\varepsilon_{\rm tot} = 1/2 u^2 + \varepsilon_{\rm th}$
is the sum of the specific internal
($\varepsilon_{\rm th} = c_{\rm v} T$) and kinetic energy.
$P$ is the vertically integrated (two-dimensional) pressure
\[
       P = \frac{R \Sigma T}{\mu}
\]
with the midplane temperature $T$ and the mean molecular weight $\mu$,
which is obtained by solving the Saha rate equations for Hydrogen dissociation
and ionization and Helium ionization.

The gravitational potential $\Phi$, generated by the binary stars,
is given by
\beq
 \label{eq:potential}
 \Phi = -\sum_{i=1,2}{\left\{\begin{array}{l@{\quad}l}
\frac{\displaystyle G\,M_i}{\displaystyle 
  |{\vec{r}} - {\vec{r}_i}|} & 
     {\rm for} |{\vec{r}} - {\vec{r}_i}|>R_* \\[1em]
\frac{\displaystyle G\,M_i\left(3R_*^2
    -|{\vec{r}} - {\vec{r}_i}|^2\right)}
     {\displaystyle 2 \, R_*^3} & 
   {\rm for} |{\vec{r}} - {\vec{r}_i}| \le R_*
\end{array}\right.}
\eeq
where $\vec{r}_i$
are the radius vectors to the two stars, and $R_*$ is the stellar radius
assumed to be identical for both stars.
The second case in Eq.~(\ref{eq:potential}) gives the potential inside
a star with radius $R_\star$ having a homogeneous (constant) density.
\subsection{Viscosity and Disk Height}
The effects of viscosity are contained in the viscous stress tensor
${\mathbf T}$. Here we assume that the accretion disk may be described
as a viscous medium driven by some internal turbulence which
we approximate with a Reynolds ansatz for the stress tensor.
The components of ${\mathbf T}$
in different coordinate systems are spelled out explicitly for
example in \citet{1978trs..book.....T}.
A useful form for disk calculations considering angular momentum
conservation is given in \citet{1999MNRAS.303..696K}.

For the kinematic shear viscosity $\nu$ we assume in this work
an $\alpha$-model \citep{1973A&A....24..337S} in the form
\beq
      \nu = \alpha c_{\rm s} H
\eeq
where $c_{\rm s}$ and $H$ are the local sound-speed and vertical height,
respectively. The local vertical height $H(\vec{r})$ is computed from
the vertical hydrostatic equilibrium
\beq
      \pdoverd{\Phi}{z} = \frac{1}{\rho} \pdoverd{p}{z} 
\eeq
where $\rho$ and $p$ are the standard three dimensional density and
pressure.
Substituting Eq.~(\ref{eq:potential}) for the potential with
$|{\vec{r}} - {\vec{r}_i}|>R_*$, assuming an ideal
gas law $p= R \rho T / \mu$, and integrating over $z$ yields
\beq
         \rho = \rho_0 e^{-\left( \frac{1}{2} \frac{z^2}{H^2} \right)}
\eeq
where $\rho_0 = \Sigma / [ (2 \pi)^{1/2} H ]$
is the midplane density, and $H$ is the vertical height
given by
\beq
  \label{eq:diskheight}
  H({\vec{r}}) = 
  \left(\sum_{i=1,2}{\frac{G M_i}
     {c_{\rm s}^2\left|{\vec{r}} - {\vec{r}_i}\right|^3}}
  \right)^{-\frac{1}{2}}
   =  \left(\sum_{i=1,2}{H_i^{-2} (\vec{r})}
  \right)^{-\frac{1}{2}}
\eeq
which can be split into the single star disk heights given
by $H_i({\vec{r}})
=c_{\rm s}\sqrt{\frac{\left|{\vec{r}}-{\vec{r}_i}\right|^{3}}{G M_i}}$.
The sound speed is given here by $c_{\rm s} = R T / \mu$.
We assume a vanishing physical bulk viscosity $\zeta$, but consider
it in the artificial viscosity coefficient
\citep[see][]{1999MNRAS.303..696K}.
\subsection{Radiative Balance}
The influence of radiative and viscous effects on the disk
temperature are treated in a local balance.
The balance equation for the internal and radiative energy considering
only these two effects reads
\beq
  \label{eq:energybalance}
  \pdoverdt{\left(\Sigma c_{\rm v} T + a T^4\right)} 
       = Q_{\rm diss} - Q_{\rm rad}
\eeq
where $Q_{\rm diss}$ and $Q_{\rm rad}$ denote the viscous dissipation
and radiative losses, respectively.
For the viscous losses we use the vertically averaged expression
\beq
     Q_{\rm diss} = {\vec{u} \cdot \nabla {\mathbf T}} 
    = \frac {1}{2 \Sigma \nu} Tr \left({\mathbf T}^2 \right)
\eeq
For the radiative transport
\beq
        Q_{\rm rad} = - \nabla \cdot {\vec{F}_0} - \pdoverd{F_z}{z}
\eeq
where $\vec{F}_0$ is the flux vector in the $z=0$ plane.
Here we consider only the losses in the vertical direction,
i.e. $\vec{F}_0 =0$, a standard approximation in accretion disk theory.
Integration over the vertical direction yields
\beq
        Q_{\rm rad} = 2 F_{\rm rad} = 2 \sigma_{\rm B} T^4_{\rm eff}
\eeq
with the local effective (surface) temperature $T_{\rm eff}$
which is related to the midplane temperature $T$ through
\beq
       T_{\rm eff}  =  \tau^{1/4} T  =  
         \left( \frac{1}{2} + \frac{3}{4} \kappa \Sigma \right)^{-1/4} T
\eeq
using an interpolated opacity $\kappa(\rho_0, T)$ adapted from
\citet{1985prpl.conf..981L}.
Note that this formulation of radiative transfer is a very good approximation
only in the optically thick regime.

\section{Numerical Issues}
\label{sec:NM}
In order to study the disk evolution, 
we utilize a finite difference method to solve the hydrodynamic equations 
outlined in the previous section.
As we intend to model possible accretion flow onto the binary
stars and to achieve a very good resolution in their vicinity,
we use a {\it Dual-Grid}-technique.

\subsection{General}
\label{subsec:general}
The code is based on the hydrodynamics code {\it RH2D}
suited to study general two-dimensional systems including
radiative transport \citep{1989A&A...208...98K}.
\textsc{RH2D} uses a spatially second-order accurate, mixed explicit 
and implicit method.
Due to an operator-splitting method it is semi-second order
accurate in time.
The advection is computed by means of the second order monotonic
transport algorithm, introduced by \citet{1977JCoPh..23..276V},
which guarantees global conservation of mass and angular momentum.
Advection and forces are solved explicitly, and the
viscosity and radiation are treated implicitly.
The formalism for application to thin disks in $(r -\varphi)$ geometry
has been described in detail in 
\citep{1998A&A...338L..37K, 1999MNRAS.303..696K}.
Here this scheme is extended to allow for a Dual-Grid system 
(see \ref{subsec:dgt}) and a more detailed
energy balance Eq.~(\ref{eq:energybalance}).

As the calculations have shown, the system behaves extremely dynamically
with very strong, moving gradients at the inner edge of the disk. To
model such a situation, the viscosity has to be treated implicitly to
ensure numerical stability.
The coupling of $u_r$ and $u_\varphi$ requires a solution method
similar to that in \citet{1989A&A...208...98K}.
For stability purposes an additional implicit artificial viscosity has to be included in the computations \citep{1999MNRAS.303..696K}.
 
This code has been employed successfully in related simulations
dealing with embedded planets in disks 
\citep{1999MNRAS.303..696K, 2000MNRAS.313L..47K}.
The numerical details of the present simulations are presented
in \citet{richi2001}.
\subsection{Dual-Grid technique}
\label{subsec:dgt}
The {\it Dual-Grid} technique employed here combines two desirable
features of thin disk computations: a) An ideally suited coordinate system
$(r-\varphi)$ in the outer bulk parts of the disk which ensures conservation
of angular momentum, and b) Coverage of the region near the central binary
star allowing the determination of the mass flow onto the stars.

In the main part the domain is covered by a cylindrical grid.
On this (outer) grid the equations are formulated in
polar coordinates, and the solution technique guarantees
exact conservation of global angular momentum. The conservation
property is absolutely crucial in obtaining physically 
reliable results for long term disk evolution 
\citep{1998A&A...338L..37K}.

\begin{figure}
\begin{center}
\resizebox{0.9\linewidth}{!}{%
\includegraphics{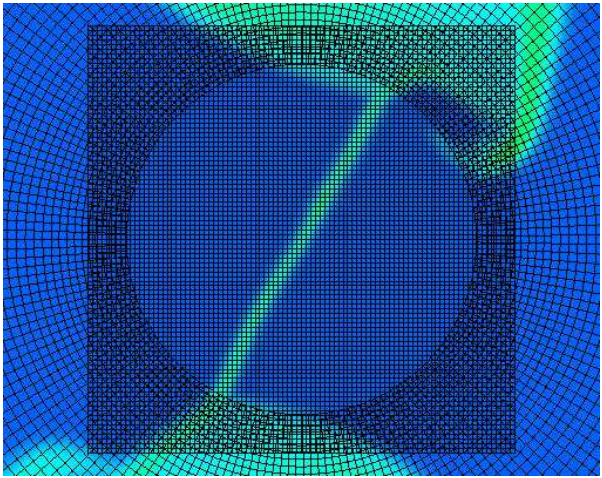}}
\end{center}
\caption{Grid system of a typical calculation. Shown
is the grid structure of the Cartesian and cylindrical grid at the
center of the computational domain. Circumstellar disks of the system
can be seen at the top and bottom edges of the picture.}
\label{fig:grids}
\end{figure}

However, such a grid cannot be extended to the very center and
leaves a hole in the middle. For typical computations
this does not pose serious problems, but in this case we
are dealing with eccentric binary systems in the middle. Hence, the
stars or their circumstellar disks may easily cross the inner grid boundary causing strong
irregularities. In any case a central hole does not allow for an accurate
determination of the individual mass accretion rates onto or a possible
overflow between the stars. 
Therefore, using a Cartesian grid centered on the origin enables an accurate
calculation of the mass flows in the vicinity of the stars.
Applying this technique, a high local
resolution can be obtained in the center as well.
An example of such a grid structure is displayed in Fig.~\ref{fig:grids}.

Related numerical schemes using a nested grid technique, though not
in different coordinate systems, have been
applied in astrophysical simulations by a number of authors
\citep[eg.][]{1992A&A...265...82R, 1993ApJ...411..274Y}.

The necessary equations (in Cartesian and cylindrical coordinates)
are then integrated, {\em independently} and {\em in parallel}, on the two grids, and
after each timestep the
necessary information in the overlap region has to be exchanged.
Restrictions are imposed on the time step 
for stability reasons, the Courant-Friedrichs-Lewy 
(CFL) condition must be fulfilled during each integration,
on each grid.

Test calculations using a single Cartesian grid covering the whole
domain have shown that the non-conservation of
angular momentum yields serious deviations from known analytical
results in particular for longer integration times.

As the grids possess a different geometry, exact conservation
of momenta and energy cannot be guaranteed in the
conversion/interpolation process which uses
linear interpolation to exchange density, temperature and velocity.
Exchanging density, internal energy and momentum was also implemented
and leads to more correct conservation of momenta and energy.
However, the advantage of a complete coverage of the physical
space in between the two stars clearly outweighs this
disadvantage.
\subsection{Tests}
We ran several tests to check the accuracy and numerical stability
of the grid overlap
region, including an infalling shock and a Sedov like explosion. For
these tests
no viscosity nor radiative effects are applied.
All tests show that the grid overlap technique is sufficient
for handling problems with radial symmetric features such as accretion
disks.

\begin{figure}
\begin{center}
\resizebox{0.9\linewidth}{!}{%
\includegraphics{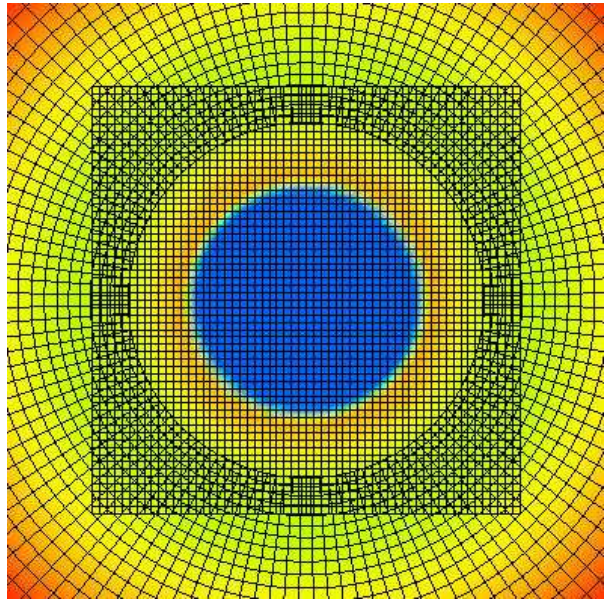}}
\end{center}
\caption{Radial shock in first (infalling) phase right after crossing the grid boundaries. Surface density is color coded, the grid structure is shown.}
\label{fig:Shock-40-g}
\end{figure}

\begin{figure}
\begin{center}
\resizebox{0.9\linewidth}{!}{%
\includegraphics{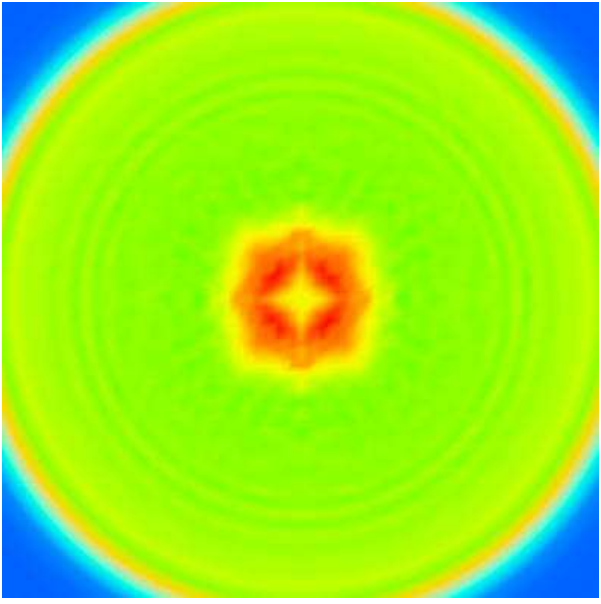}}
\end{center}
\caption{Radial shock in second (explosion) phase right after crossing the grid boundaries again. Surface density is color coded, the imaged domain is the same as in Fig.~\ref{fig:Shock-40-g}.}
\label{fig:Shock-160}
\end{figure}

The shock experiment initial setup is a step function of density
and temperature with the floor starting at $r_0$ (about 10 grid cells outside of the inner boundary of the cylindrical grid) and covering the whole Cartesian grid:
\beq
\label{eq:shock}
  \Sigma(r) = \Sigma_{\rm min} + \Sigma_0\,\Theta(r-r_0)\,.
\eeq
In Fig.~\ref{fig:Shock-40-g} you can see the grid setup and the shock in
the initial infalling phase just after crossing the grid boundaries. The shock
front remains radially symmetric and no reflections are visible. After
the shock compresses into a single point it switches to the outflowing
phase. You can see the shock front propagating outwards just after crossing
the grid boundaries in Fig.~\ref{fig:Shock-160}.
Apart from unavoidable grid effects near the center the shock front remains nicely circular and again neither disturbances nor reflections are visible at the grid interface.

\section{General model design}
\label{sec:GMD}
The main goal of this study is the investigation of the characteristic
features of the hydrodynamic flow of the circumbinary disk.

From now on, we refer to non-dimensional units.
All the lengths are expressed in units of the semi-major
axis $a$ of the binary.
This is constant because the time scale on which the binary
orbital elements change due to mass and angular momentum transfer 
is much longer than the dynamical time scales considered here
\citep[see][]{2001IAUS..200..439A}.
Masses are in units of the solar mass and
time is given in units of the binary's orbital period. 

The whole azimuthal range of the disk is taken into account
by considering a computational domain represented by
$2\,\pi \times [0, r\subscr{out}]$, where
typically $r\subscr{out}=10$ in units of the binary separation.
The cylindrical grid covers a region $2\,\pi \times [r\subscr{in}, r\subscr{out}]$ with $r\subscr{in}=0.1$ represented by up to $256 \times 256$ meshpoints, and
the central Cartesian grid has up to $128 \times 128$ meshpoints covering
$2\,r\subscr{in} \times 2\,r\subscr{in}$ with an additional small overlap.
\subsection{Accretion onto the Stars}
\label{subsec:accret}
As will be seen, matter flows from the disk through the gap and
circulates the stars in small circumstellar disks. From those it can
be accreted onto the stars.  This stellar accretion is accounted for
by removing some mass from a region close to the stars.  This
removal is performed on both grids dependent on the star positions.
The removed mass is not added to the dynamical mass of the stars, but
is just monitored.

The region $R$ out of which matter is removed is given as a fraction
(0.2) of the Roche lobe size for the wide systems and as the stellar
cores for the tight spectroscopic systems. The accretion process typically involves
only a few grid cells on each side of the stars for the wide systems and
the whole resolved stellar core (up to 11x11 grid-cells) for the tight
systems, making it a locally confined process.

Two methods for deciding whether a grid cell is targeted for accreting
a part of its mass were implemented. Accreting a constant
fraction $f$ per time of the mass exceeding a defined minimum density $\Sigma_{\rm min}$ leads to a very smooth accretion process.
The accreted mass per iteration is computed by
\beq
\label{eq:accreting1}
  M_{\rm acc} = \sum_{i,j \in R} {\rm min}(1.0,f\Delta t)\, {\rm max}(0.0, \Sigma_{ij} - \Sigma_{\rm min}) \Delta{\rm Vol}_{ij}
\eeq
Limiting the density in the accretion region $R$ to an arbitrary
value $\Sigma_{\rm max}$ results in a more realistic structure of the circumstellar disks
and to more correct spectral energy distributions of the circumstellar
disks.
\beq
\label{eq:accreting2}
  M_{\rm acc} = \sum_{i,j \in R} {\rm max}(0.0, \Sigma_{ij} - \Sigma_{\rm max})\,\Delta{\rm Vol}_{ij}
\eeq
This accretion process is not smooth, but consists of accretion
events because exceeding the maximum density is a discrete process
and not distributed evenly over time.
Nevertheless the time-averaged accretion rates are the same for both
methods.

The inferred accretion rates in any event compare favorably with the average
disk accretion rate, as estimated from the radial mass fluxes through different
surfaces, and the evolution of the
total mass found in the circumstellar disks.

\subsection{{\bf Spectra generation}}

Spectral energy distributions are generated decoupled from the
dynamic evolution of the disk as a post-processing step. Here the
model from \citet{1988ApJ...326..865A} is extended to work on
data generated by our numerical simulation. The flux at the observer
is obtained by integrating over the disk surface assuming local
black-body radiation
\beq
  \label{eq:sed}
  F_\nu = \frac{\cos i}{D^2}\int_{r_0}^{R_d}\int_{0}^{2\pi}B_\nu\left[T(r,\varphi)\right]\left(1-e^{-\tau(r,\varphi)}\right)\,r{\rm d}\varphi\,{\rm d}r
\eeq
where $i$ is the inclination of the disk, $D$ the distance to the observer
and $\tau$ the line-of-sight optical depth through the disk. $\tau$ can
be obtained using the opacity $\kappa$, $\tau(r,\varphi)=\frac{\kappa \Sigma(r,\varphi)}{\cos i}$. $T$ is the surface disk temperature which has to
be estimated using the optical depth of the disk obtained from an opacity table.

The integration is limited to regions outside of the stellar cores as
temperatures and densities from inside the cores can be nothing but
wrong. The star emission is accounted for by adding a black-body spectrum
from an appropriate effective star temperature.

\subsection{Initial and boundary conditions}
\label{subsec:initial}
\begin{figure}
\begin{center}
\resizebox{0.9\linewidth}{!}{%
\includegraphics{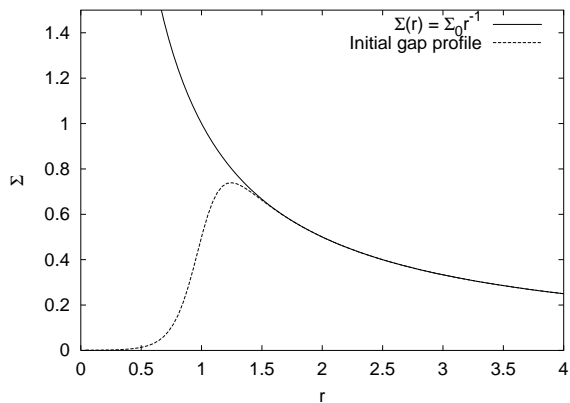}}
\end{center}
\caption{Initial gap profile composed out of $\Sigma_0 r^{-1}$ and
the gap function Eq.~(\ref{eq:gap}).
The figure shows a gap at $r\subscr{gap}=1$ with $\Sigma_0=1$.}
\label{fig:gap-profile}
\end{figure}

The initial density distribution is proportional to $r^{-d}$, where
$d$ is either set to $0.5$ which is the equilibrium density distribution
for an accretion disk around a central star having a constant kinematic viscosity, or given by a fit of the
generated spectral energy distribution to the one observed.
However, we superimpose to this an axisymmetric gap around the 
COM of the two stars,
given by an approximate gap radius $r_{\rm gap}$ \citep{1994ApJ...421..651A}
and a gap function
\beq
\label{eq:gap}
  f_{\rm gap} := \left(e^{-\frac{r-r\subscr{gap}}{0.1r\subscr{gap}}}+1\right)^{-1}\,.
\eeq
Figure~\ref{fig:gap-profile} shows a typical surface density at $t=0$.
The initial velocity field is that of a Keplerian disk orbiting
the center of mass of the two stars.

The gravitational potential in Eq.~(\ref{eq:potential})
is phased-in smoothly from an initial central potential for the
total mass, $M_1 + M_2$,
during an initialization phase lasting typically ten orbital
periods. This allows the system to adjust smoothly from the semi-stationary
setup of a single central star to the binary star situation.

For the boundary conditions, periodicity is imposed at
$\varphi=0$ and $\varphi=2\,\pi$.
We set outflowing boundary 
conditions at the outer radial border ($r\subscr{out}$), i.e.
the radial velocity gradient at $R_{\rm max}$ is zero, and truncated to
positive values.
Angular velocity there equals the unperturbed initial Keplerian value.

For single grid calculations with a cylindrical grid we use zero
gradient boundary conditions for both velocity components and the
density at the inner border.
\section{Results}
\label{sec:results}
Before discussing the results concerning particular astronomical
objects, we would like to present the details of a specific
test case of an equal mass binary star on a circular orbit.
Such a system is well suited to check important properties
(symmetries, stability) of the numerical method used. Also general
properties of accretion disks in binary systems can be derived from
such a test case.
\subsection{Equal Mass Binary}
\label{subsec:emb}
The equal mass binary on a circular orbit serves as a testing platform for numerical stability and accuracy. In a coordinate frame corotating with the binary one expects after some transient time a quasi-stationary situation, which is symmetric with respect to the line connecting the stars. As the viscosity is solved implicitly by solving a large matrix system iteratively, exact symmetry cannot be preserved numerically. Also usual density contrasts are of the order of $10^3$ for numerical floor to circumstellar disks and from circumstellar disks to the circumbinary disk. Hence, this test serves as an excellent example for estimating the ability to preserve symmetry over long time scales.

\begin{figure}
\begin{center}
\resizebox{0.9\linewidth}{!}{%
\includegraphics{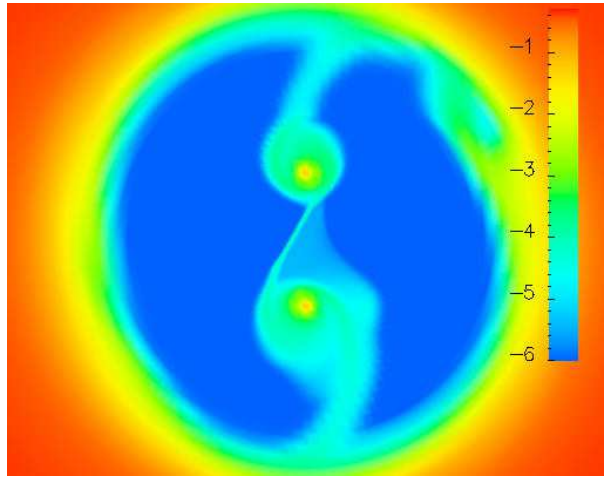}}
\end{center}
\caption{Evolution of an equal mass binary after 90 orbital periods.
Color encoding is logarithmic surface density.
Both the inner part of the circumbinary disk and the circumstellar
disks are visible.}
\label{fig:reference}
\end{figure}

In Fig.~\ref{fig:reference} you can see a circumbinary disk after 90 orbits evolution time in almost perfect quasi-stationary state. The deviation from line-symmetry is small. Only with very high resolution simulations we observe spiral features in the outer regions of the disk, while spiral arms originating from the inner gap of the circumbinary disk flowing onto the circumstellar disks can be observed even for low resolutions. Those spiral arms feeding the circumstellar disks remain stable for the whole period of a non-eccentric binary.

The accretion rates for both accretion mechanisms onto the two stars are nearly identical and do not show any significant variations for different numerical parameters. Thus, our implementation of the accretion mechanism is robust. 

\begin{figure}
\begin{center}
\resizebox{0.9\linewidth}{!}{%
\includegraphics{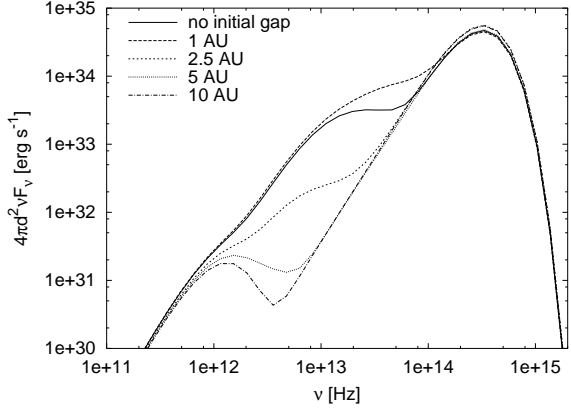}}
\end{center}
\caption{Dependency of the gap width on the SED.}
\label{fig:sed-gap}
\end{figure}

\begin{figure}
\begin{center}
\resizebox{0.9\linewidth}{!}{%
\includegraphics{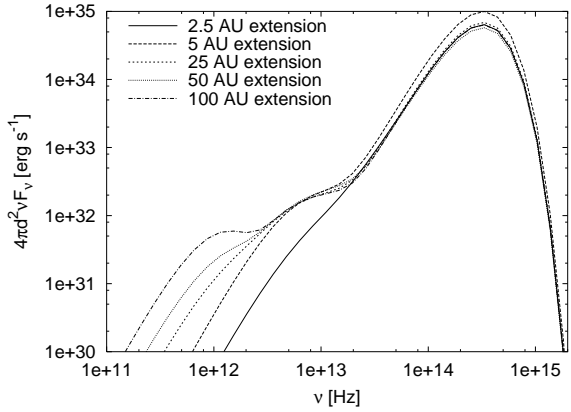}}
\end{center}
\caption{Dependency of the disk size on the SED.}
\label{fig:sed-size}
\end{figure}

\begin{figure}
\begin{center}
\resizebox{0.9\linewidth}{!}{%
\includegraphics{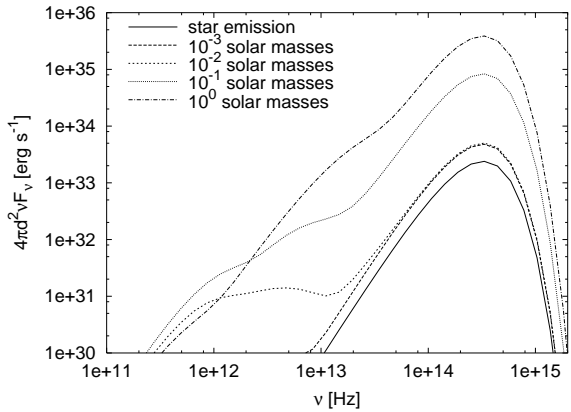}}
\end{center}
\caption{Dependency of $\Sigma_0$ on the SED.}
\label{fig:sed-sigma0}
\end{figure}

Changing model parameters such as $\Sigma_0$, gap width, or disk size we
obtain characteristic variations of the resulting spectral energy
distribution of an evolved disk. Figures~\ref{fig:sed-gap}-\ref{fig:sed-sigma0}
show dependencies of these parameters on the spectral energy distribution
that are used to fit parameters of the spectroscopic systems to match
observed spectra.

\subsection{Spectroscopic Systems}
\begin{figure}
\begin{center}
\resizebox{0.9\linewidth}{!}{%
\includegraphics{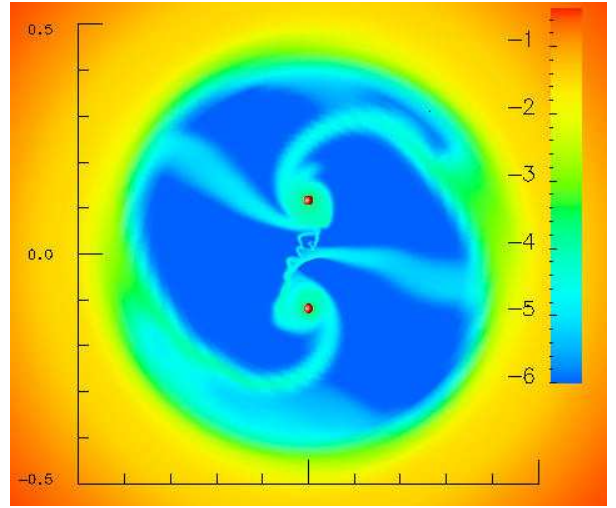}}
\end{center}
\caption{\object{AK Sco} circumbinary disk after 82.0 orbital periods in apastron. Color coding is $\log(\Sigma)$, the size of the stars reflects the actual stellar radii, the length scales are in AU}
\label{fig:AK-Sco-pic}
\end{figure}

\begin{figure}
\begin{center}
\resizebox{0.9\linewidth}{!}{%
\includegraphics{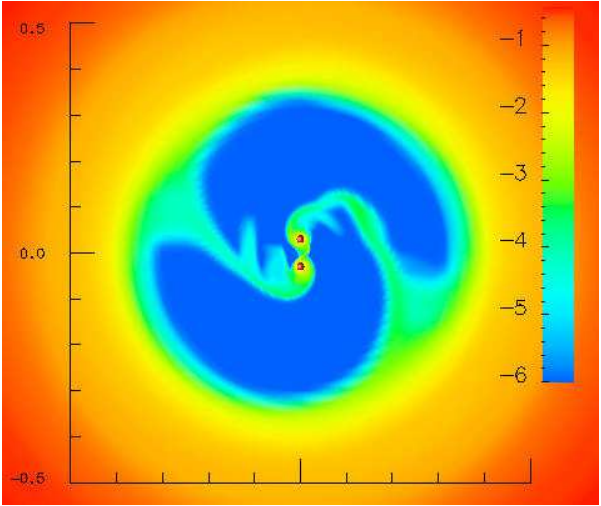}}
\end{center}
\caption{\object{DQ Tau} circumbinary disk after 85.5 orbital periods in periastron. Color coding is $\log(\Sigma)$, the size of the stars reflects the actual stellar radii, the length scales are in AU.}
\label{fig:DQ-Tau-pic}
\end{figure}

Hereafter we discuss models for the two spectroscopic
young binary stars \object{DQ Tau} and \object{AK Sco} which 
have a separation of only a fraction of an $\AU$.
The physical parameter for the systems are taken from
\citet{1989A&A...219..142A,1998A&A...335..218F,1997AJ....114..301J}
for \object{AK Sco} and \citet{1997AJ....113.1841M}
for \object{DQ Tau}.
The relevant information for the
simulations has been summarized in Table \ref{table:spectroscopic},
disk masses $M_{\rm d}$ are integrated densities over the computational
domain specified by the disk radii $R_{\rm d}$.
Images showing the disk configuration in apastron and periastron
phase are shown in Figs.~\ref{fig:AK-Sco-pic} and \ref{fig:DQ-Tau-pic}.

Generally, for eccentric systems we observe spiral features in the outer
circumbinary disk only for very high resolution simulations, while again
spiral arms feed the circumstellar disks even with low resolutions. Opposed
to the non-eccentric systems the spiral arms develop during apastron phase
and are torn off together with the outer parts of the circumstellar disks
after periastron phase for high eccentric systems. This tearing off is caused
by raising centrifugal forces acting on the circumstellar material.

\begin{table}
\begin{center}
\begin{tabular}{l|l|l}
               & AK Scorpii           & DQ Tauri \\ \hline\hline
$P$            & $13^d.60933\pm0^d.00040$ & $15^d.8043\pm0^d.0024$ \\ \hline
$e$            & $0.469\pm0.004$ & $0.556\pm0.018$ \\ \hline
$q$            & $0.987\pm0.007$ & $0.97\pm0.15$ \\ \hline
$i$            & $63^\circ$ & $23^\circ$ \\ \hline
$a$            & $0.16\,\AU$ & $0.135\,\AU$ \\ \hline
$T_*$          & 6500\,K & 4000\,K \\ \hline
$R_*$          & $2.2\,\RSOL$ & $1.785\,\RSOL$ \\ \hline
$M_*$          & $1.5\,\MSOL$ & $1.3\,\MSOL$ \\ \hline
$d$            & $152\,\PC$ & $140\,\PC$ \\ \hline
$M_{\rm d}$    & $0.005\,\MSOL$ & $0.001\,\MSOL$ \\ \hline
$R_{\rm d}$    & $40\,\AU$      & $50\,\AU$ \\
\end{tabular}
\end{center}
\caption{Parameters for \object{AK Sco} and \object{DQ Tau}.}
\label{table:spectroscopic}
\end{table}

For both systems we compared the accretion rates resulting from the
simulation with accretion rates derived from of observations as done
by \citet{1995ApJ...452..736H} and \citet{1998ApJ...492..323G}
(see Table~\ref{table:rates}). As no data was available for \object{AK Sco},
a comparison with observational data was not possible, but as both
spectroscopic systems have similar system parameters, accretion rates
of the same order of magnitude can be expected.

\begin{table}
\begin{center}
\begin{tabular}{|l|l|l|l|l|l|}\hline
       & Hartigan & Gullbring & simulation \\ \hline
DQ Tau & $0.50\cdot 10^{-7}$ & $0.60\cdot 10^{-9}$  & $0.62\cdot 10^{-8}$ \\ \hline
AK Sco & n.a.              & n.a.                 & $0.83\cdot 10^{-8}$ \\ \hline
GG Tau & $0.20\cdot 10^{-6}$ & $0.175\cdot 10^{-7}$ & $0.11\cdot 10^{-8}$ \\ \hline
UY Aur & $0.25\cdot 10^{-6}$ & $0.656\cdot 10^{-7}$ & $0.5\cdot 10^{-7}$ \\ \hline
\end{tabular}
\end{center}
\caption{Accretion rates in $\MSOL{\rm yr}^{-1}$ derived from observational data from \citet{1995ApJ...452..736H} and \citet{1998ApJ...492..323G} compared to rates as resulting from our simulations. Our rates are for the primary component of the binaries, for the other data we assume it is an averaged accretion rate for both stars.}
\label{table:rates}
\end{table}

\begin{figure}
\begin{center}
\resizebox{0.9\linewidth}{!}{%
\includegraphics{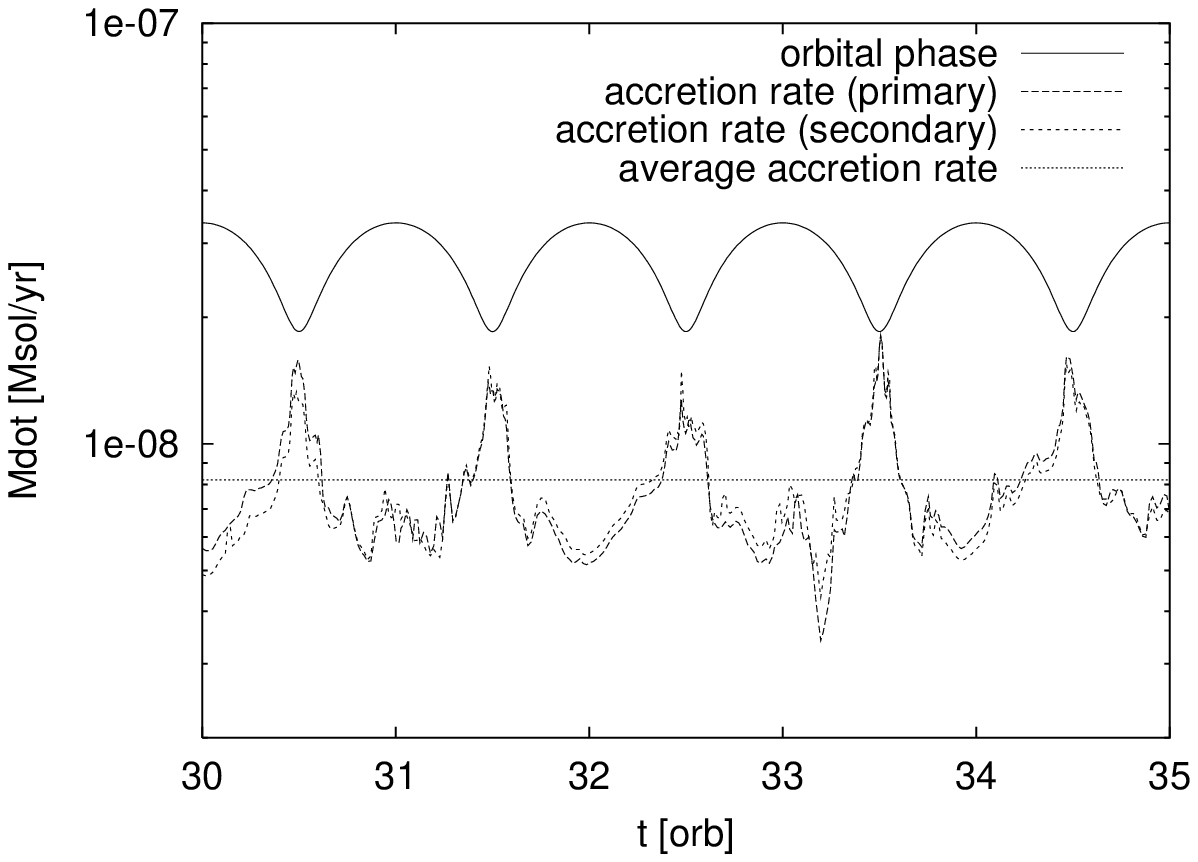}}
\end{center}
\caption{Time-dependent accretion rate of the \object{AK Sco} system. The orbital phase curve schematically shows the distance of the two stars.}
\label{fig:AK-Sco-acc-detail}
\end{figure}

\begin{figure}
\begin{center}
\resizebox{0.9\linewidth}{!}{%
\includegraphics{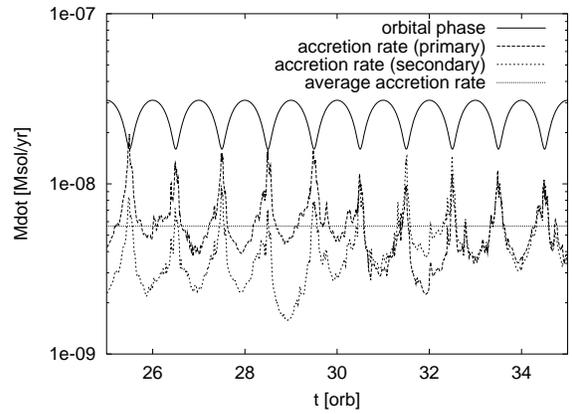}}
\end{center}
\caption{Time-dependent accretion rate of the \object{DQ Tau} system. The orbital phase curve schematically shows the distance of the two stars.}
\label{fig:DQ-Tau-acc-detail}
\end{figure}

As can be seen from Figs.~\ref{fig:AK-Sco-acc-detail} and \ref{fig:DQ-Tau-acc-detail} the accretion process happens synchronized with periastron phase of the binaries. This is in agreement with numerical simulations from \citet{1996ApJ...467L..77A} and \citet{1997apro.conf..792R} and with observations which show excess luminosity at these phases \citep{1997AJ....113.1841M}.

From our simulations we can derive an averaged accretion rate of $0.83\cdot 10^{-8}\,\MSOL{\rm yr}^{-1}$ for the primary, $0.82\cdot 10^{-8}\,\MSOL{\rm yr}^{-1}$ for the secondary of \object{AK Sco} and $0.62\cdot 10^{-8}\,\MSOL{\rm yr}^{-1}$ for the primary, $0.51\cdot 10^{-8}\,\MSOL{\rm yr}^{-1}$ for the secondary of \object{DQ Tau}. These accretion rates are in good agreement with the observational rates.

It is expected that the luminosity increases in periastron because tidal compression of the disk leads to a temperature rise with subsequent luminosity increase.
We tried to generate luminosity curves to show that indeed excess luminosity is produced at periastron phase, but unfortunately those curves show both brightening and darkening dependent on the implemented accretion mechanism.
At the very low averaged accretion rates of $\dot M \simeq 7\cdot10^{-9}\,\MSOL{\rm yr}^{-1}$ the disk is very optically thin. In that regime the implemented radiative loss mechanism is not very accurate and thus these results are not meaningful.

\begin{figure}
\begin{center}
\resizebox{0.9\linewidth}{!}{%
\includegraphics{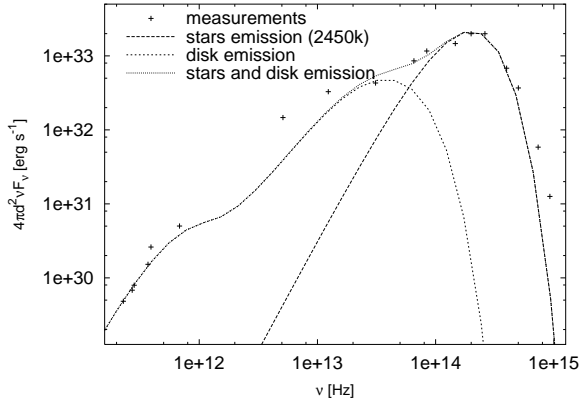}}
\end{center}
\caption{Spectral energy distribution of \object{DQ Tau} ($d=2.0$), crosses show observational data taken from \citet{1997AJ....113.1841M}.}
\label{fig:DQ-Tau-sed}
\end{figure}

\begin{figure}
\begin{center}
\resizebox{0.9\linewidth}{!}{%
\includegraphics{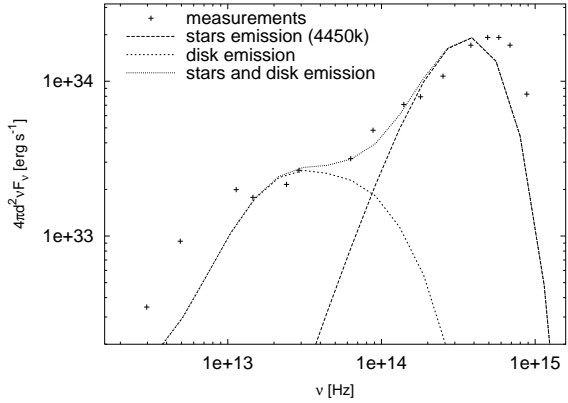}}
\end{center}
\caption{Spectral energy distribution of \object{AK Sco} ($d=1.5$), crosses show observational data taken from \citet{1997AJ....114..301J}.}
\label{fig:AK-Sco-sed}
\end{figure}

For the spectroscopic binaries we can generate spectral energy distributions out of the simulation data which can be made to match the observed ones by manually adjusting initial model parameters such as disk mass, gap width and density profile $d$. For \object{DQ Tau} Fig.~\ref{fig:DQ-Tau-sed} shows the spectral energy distribution composed out of the disk and the star, Fig.~\ref{fig:AK-Sco-sed} shows the one for the \object{AK Sco} system.
The effective stellar temperatures used in the model have been adjusted to match the observational data. They differ from the quoted values in the literature in Table~\ref{table:spectroscopic}. This may be due to difficulties in separating individual stellar and disk contributions to the total flux in models and observations.
Also occultation is not accounted for on adding the star spectrum to the disk.

\subsection{Wider Systems}
\label{subsec:wide}
There are only two well known systems where the binary
and the disk have been imaged directly,
\object{GG Tau} \citep{1994A&A...286..149D,1999A&A...348..570G}
and \object{UY Aur} \citep{1998ApJ...499..883C,1998A&A...332..867D}.
These systems have a much wider separation of tens to over 100 AU,
a mass ratio different from unity
and show consequently a quite different behavior. In Table~\ref{table:optical}
the system parameters used for the simulations are shown.

\begin{table}
\begin{center}
\begin{tabular}{l|l|l}
               & UY Aurigae  & GG Tauri \\ \hline\hline
$P$            & $2074\,$yr    & $490\,$yr \\ \hline
$e$            & $0.13$      & $0.25$ \\ \hline
$q$            & $0.63$      & $0.77$ \\ \hline
$i$            & $42^\circ$  & $43^\circ\pm 5^\circ$ \\ \hline
$a$            & $190\,\AU$    & $67\,\AU$ \\ \hline
$T_*$          & 4000\,K       & 3800\,K \\ \hline
$R_*$          & $2.6\,\RSOL$  & $2.8\,\RSOL$ \\ \hline
$M_*$          & $0.95\,\MSOL$ & $0.65\,\MSOL$ \\ \hline
$d$            & $140\,\PC$    & $150\,\PC$ \\ \hline
$M_{\rm d}$    & $1.1\,\MSOL$  & $0.15\,\MSOL$ \\ \hline
$R_{\rm d}$    & $2100\,\AU$    & $600\,\AU$ \\
\end{tabular}
\end{center}
\caption{Parameters for \object{UY Aur} and \object{GG Tau}.}
\label{table:optical}
\end{table}

From our simulations we derive averaged accretion rates for both
\object{UY Aur} and \object{GG Tau} that are comparable with the
observed quantities in Table~\ref{table:rates}.
In the case of \object{UY Aur} we get $0.5\cdot 10^{-7}\,\MSOL{\rm yr}^{-1}$ for the primary and $0.25\cdot 10^{-6}\,\MSOL{\rm yr}^{-1}$ for the secondary, respectively.
Simulations of \object{GG Tau} lead to accretion rates of
$0.11\cdot 10^{-8}\,\MSOL{\rm yr}^{-1}$ for the primary and
$0.33\cdot 10^{-9}\,\MSOL{\rm yr}^{-1}$ for the secondary. As expected from
the difference of the disk masses ($0.15\,\MSOL$ vs.~$1.1\,\MSOL$) the accretion rate for \object{GG Tau} is one order of
magnitude less than the one for \object{UY Aur}.

Accretion is reinforced at periastron phases as for the spectroscopic systems,
but events are an order of magnitude lower for \object{GG Tau} and nearly
vanish for \object{UY Aur} due to the lower eccentricity of the systems. Also both
systems show more phase dependent accretion for the secondary than for the
primary.

\begin{figure}
\begin{center}
\resizebox{0.9\linewidth}{!}{%
\includegraphics{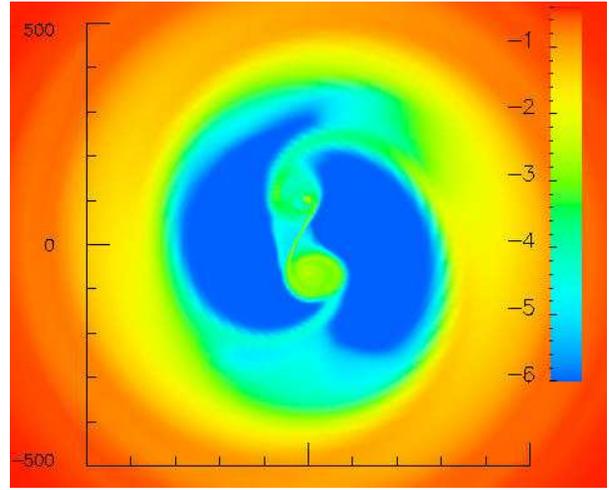}}
\end{center}
\caption{\object{UY Aur} circumbinary disk after 20.5 orbital periods. Color
coding is $\log(\Sigma)$, the length scales are in AU.}
\label{fig:UY-Aur-pic}
\end{figure}

\begin{figure}
\begin{center}
\resizebox{0.9\linewidth}{!}{%
\includegraphics{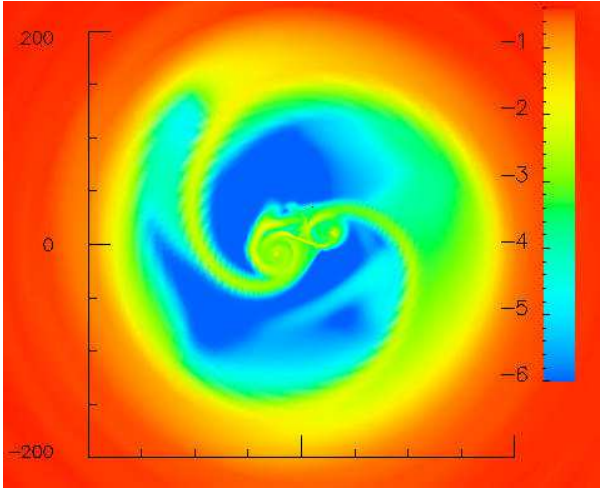}}
\end{center}
\caption{\object{GG Tau} circumbinary disk after 56.4 orbital periods. Color
coding is $\log(\Sigma)$, the length scales are in AU.}
\label{fig:GG-Tau-pic}
\end{figure}

In Figs.~\ref{fig:UY-Aur-pic} and \ref{fig:GG-Tau-pic} you can see
the inner parts of the simulated circumbinary disks. The circumstellar
disks remain intact over the whole period.
It was not possible to generate meaningful spectral
energy distributions of the disks as they are too cold for these wide
systems. Also observational data is lacking for these systems. 

\section{Conclusion}
\label{sec:conclusion}
With the present calculations we have modeled the
accretion flow onto the stars within a circumbinary disk
in more detail using more realistic physics, much higher
resolution and a much longer time integration.
Our main achievements may be summarized as follows:
\begin{enumerate}
\item  Development of a new Dual-Grid technique combining
       two coordinate systems, which is ideally suited for
       modeling planar accretion disks including their 
       central objects.
\item  Performing long-time integrations of binary stars and
       circumbinary disks, including a detailed vertical energy balance.
\item  For close binaries with separations of only a fraction of
       an AU we find a relatively large accretion rate crossing
       the gap of the order of $10^{-8}\MSOL{\rm yr}^{-1}$.
\item  For wide binaries $(a = 10-200\,\AU)$ the accretion rate is
       comparable to those of the close binaries only because of the
       large disk masses of the two systems.
\item  For a suitable variation of the initial density distribution
       $\Sigma \sim r^{-d}$ of the disk, with $d$ between $1.5$ and $2$
       the spectral energy distributions of \object{DQ Tau} and \object{AK Sco}
       could be fitted well to the observational data.
\end{enumerate}
Future models of this kind need to include a more detailed structure of the stars as well as a more detailed model of the accretion process.
The numerical resolution we have reached by now is so fine that the
individual stars in close binaries are already resolved by usually about
100 grid-cells within low-res calculations.

To obtain a detailed comparison with observed light curves in particular the phase dependence a more elaborate model of the radiative loss of the very thin material in the circumstellar disks is required. In that case 3d simulations may be necessary which are beyond present day computational resources.
\begin{acknowledgements}
Part of this work was funded by the German Science Foundation (DFG)
under SFB 382
{\it Simulation physikalischer Prozesse auf H\"ochstleistungsrechnern}. 
We would like to thank Dr. Hubert Klahr for many stimulating 
discussions during this project.
\end{acknowledgements}
%
\bibliographystyle{aa}
\bibliography{cbd}

\begin{thebibliography}{31}
\expandafter\ifx\csname natexlab\endcsname\relax\def\natexlab#1{#1}\fi

\bibitem[{{Adams} {et~al.}(1988){Adams}, {Shu}, \&
  {Lada}}]{1988ApJ...326..865A}
{Adams}, F.~C., {Shu}, F.~H., \& {Lada}, C.~J. 1988, \apj, 326, 865

\bibitem[{{Andersen} {et~al.}(1989){Andersen}, {Lindgren}, {Hazen}, \&
  {Mayor}}]{1989A&A...219..142A}
{Andersen}, J., {Lindgren}, H., {Hazen}, M.~L., \& {Mayor}, M. 1989, \aap, 219,
  142

\bibitem[{{Artymowicz} {et~al.}(1991){Artymowicz}, {Clarke}, {Lubow}, \&
  {Pringle}}]{1991ApJ...370L..35A}
{Artymowicz}, P., {Clarke}, C.~J., {Lubow}, S.~H., \& {Pringle}, J.~E. 1991,
  \apjl, 370, L35

\bibitem[{{Artymowicz} \& {Lubow}(1994)}]{1994ApJ...421..651A}
{Artymowicz}, P. \& {Lubow}, S.~H. 1994, \apj, 421, 651

\bibitem[{{Artymowicz} \& {Lubow}(1996)}]{1996ApJ...467L..77A}
---. 1996, \apjl, 467, L77

\bibitem[{{Artymowicz} \& {Lubow}(2001)}]{2001IAUS..200..439A}
{Artymowicz}, P. \& {Lubow}, S.~H. 2001, in IAU Symposium, Vol. 200, 439

\bibitem[{{Bate} \& {Bonnell}(1997)}]{1997MNRAS.285...33B}
{Bate}, M.~R. \& {Bonnell}, I.~A. 1997, \mnras, 285, 33

\bibitem[{{Close} {et~al.}(1998){Close}, {Dutrey}, {Roddier}, {Guilloteau},
  {Roddier}, {Northcott}, {Menard}, {Duvert}, {Graves}, \&
  {Potter}}]{1998ApJ...499..883C}
{Close}, L.~M., {Dutrey}, A., {Roddier}, F., {et~al.} 1998, \apj, 499, 883

\bibitem[{{Duquennoy} \& {Mayor}(1991)}]{1991A&A...248..485D}
{Duquennoy}, A. \& {Mayor}, M. 1991, \aap, 248, 485

\bibitem[{{Dutrey} {et~al.}(1994){Dutrey}, {Guilloteau}, \&
  {Simon}}]{1994A&A...286..149D}
{Dutrey}, A., {Guilloteau}, S., \& {Simon}, M. 1994, \aap, 286, 149

\bibitem[{{Duvert} {et~al.}(1998){Duvert}, {Dutrey}, {Guilloteau}, {Menard},
  {Schuster}, {Prato}, \& {Simon}}]{1998A&A...332..867D}
{Duvert}, G., {Dutrey}, A., {Guilloteau}, S., {et~al.} 1998, \aap, 332, 867

\bibitem[{{Favata} {et~al.}(1998){Favata}, {Micela}, {Sciortino}, \&
  {D'Antona}}]{1998A&A...335..218F}
{Favata}, F., {Micela}, G., {Sciortino}, S., \& {D'Antona}, F. 1998, \aap, 335,
  218

\bibitem[{{Guilloteau} {et~al.}(1999){Guilloteau}, {Dutrey}, \&
  {Simon}}]{1999A&A...348..570G}
{Guilloteau}, S., {Dutrey}, A., \& {Simon}, M. 1999, \aap, 348, 570

\bibitem[{{Gullbring} {et~al.}(1998){Gullbring}, {Hartmann}, {Briceno}, \&
  {Calvet}}]{1998ApJ...492..323G}
{Gullbring}, E., {Hartmann}, L., {Briceno}, C., \& {Calvet}, N. 1998, \apj,
  492, 323

\bibitem[{{G\"unther}(2001)}]{richi2001}
{G\"unther}, R. 2001, {Dynamik und Entwicklung von zirkumbin\"aren Scheiben}
  (Diploma Thesis, University of T\"ubingen,), 64

\bibitem[{{Hartigan} {et~al.}(1995){Hartigan}, {Edwards}, \&
  {Ghandour}}]{1995ApJ...452..736H}
{Hartigan}, P., {Edwards}, S., \& {Ghandour}, L. 1995, \apj, 452, 736

\bibitem[{{Jensen} \& {Mathieu}(1997)}]{1997AJ....114..301J}
{Jensen}, E. L.~N. \& {Mathieu}, R.~D. 1997, \aj, 114, 301

\bibitem[{{Kley}(1989)}]{1989A&A...208...98K}
{Kley}, W. 1989, \aap, 208, 98

\bibitem[{{Kley}(1998)}]{1998A&A...338L..37K}
---. 1998, \aap, 338, L37

\bibitem[{{Kley}(1999)}]{1999MNRAS.303..696K}
---. 1999, \mnras, 303, 696

\bibitem[{{Kley}(2000)}]{2000MNRAS.313L..47K}
---. 2000, \mnras, 313, L47

\bibitem[{{Lin} \& {Papaloizou}(1985)}]{1985prpl.conf..981L}
{Lin}, D.~N.~C. \& {Papaloizou}, J. 1985, in Protostars and Planets II, 981

\bibitem[{{Mathieu} {et~al.}(2000){Mathieu}, {Ghez}, {Jensen}, \&
  {Simon}}]{2000prpl.conf..703M}
{Mathieu}, R.~D., {Ghez}, A.~M., {Jensen}, E.~L.~N., \& {Simon}, M. 2000,
  Protostars and Planets IV, 703

\bibitem[{{Mathieu} {et~al.}(1997){Mathieu}, {Stassun}, {Basri}, {Jensen},
  {Johns-Krull}, {Valenti}, \& {Hartmann}}]{1997AJ....113.1841M}
{Mathieu}, R.~D., {Stassun}, K., {Basri}, G., {et~al.} 1997, \aj, 113, 1841

\bibitem[{{Rozyczka} \& {Laughlin}(1997)}]{1997apro.conf..792R}
{Rozyczka}, M. \& {Laughlin}, G. 1997, in ASP Conf. Ser. 121: IAU Colloq. 163:
  Accretion Phenomena and Related Outflows, 792

\bibitem[{{Ruffert}(1992)}]{1992A&A...265...82R}
{Ruffert}, M. 1992, \aap, 265, 82

\bibitem[{{Shakura} \& {Sunyaev}(1973)}]{1973A&A....24..337S}
{Shakura}, N.~I. \& {Sunyaev}, R.~A. 1973, \aap, 24, 337

\bibitem[{{Tassoul}(1978)}]{1978trs..book.....T}
{Tassoul}, J. 1978, {Theory of rotating stars} (Princeton Series in
  Astrophysics, Princeton: University Press, 1978)

\bibitem[{{van Leer}(1977)}]{1977JCoPh..23..276V}
{van Leer}, B. 1977, Journal of Computational Physics, 23, 276

\bibitem[{{Yorke} {et~al.}(1993){Yorke}, {Bodenheimer}, \&
  {Laughlin}}]{1993ApJ...411..274Y}
{Yorke}, H.~W., {Bodenheimer}, P., \& {Laughlin}, G. 1993, \apj, 411, 274

\bibitem[{{Zinnecker} \& {Mathieu}(2001)}]{2001IAUS..200.....H}
{Zinnecker}, H. \& {Mathieu}, R. 2001, in IAU Symposium, Vol. 200

\end{thebibliography}
\end{document}